\documentstyle[12pt]{article}
\begin{document}

\title{New considerations on the separability of very noisy mixed states and implications
for NMR quantum computing}
\author{J.D. Bulnes(a), R.S. Sarthour(a), E.R. de Azevedo(b), F.A. Bonk(b), \\
 J.C.C. Freitas(c), A.P. Guimar\~aes(a), T.J. Bonagamba(b), I.S. Oliveira(a) \\
\normalsize {(a) Centro Brasileiro de Pesquisas F\'{\i}sicas} \\
\normalsize { Rua Dr. Xavier Sigaud 150, Rio de Janeiro - 22290-180, Brazil} \\
\normalsize {(b) Instituto de F\'{\i}sica de S\~ao Carlos} \\
\normalsize { Universidade de S\~ao Paulo, S\~ao Paulo - P.O. Box 369, 13560-970,  Brazil} \\
\normalsize {(c)Depto. F\'{\i}sica, Universidade Federal do Esp\'{\i}rito Santo} \\
\normalsize {Vit\'oria - 22060-900,  Esp\'{\i}rito Santo, Brazil}
} \maketitle
\begin{abstract}
\noindent We revise the problem first addressed by Braunstein and
co-workers (Phys. Rev. Lett. {\bf 83} (5) (1999) 1054) concerning 
the separability of very noisy mixed
states represented by general density matrices with the form
$\rho_\epsilon = (1-\epsilon)M_d+\epsilon\rho_1$. From a detailed
numerical analysis, it is shown that: (1) there exist infinite values
in the interval taken for the density matrix expansion
coefficients, $-1\le c_{\alpha_1,\ldots,\alpha_N}\le 1$, which
give rise to {\em non-physical density matrices}, with trace equal
to 1, but at least one {\em negative} eigenvalue; (2) there exist 
entangled matrices outside the predicted entanglement region, and (3) 
there exist separable matrices inside the same region. It is also 
shown that the lower and upper bounds of $\epsilon$ depend 
on the coefficients of the expansion of $\rho_1$ in the Pauli basis. If
$\rho_{1}$ is hermitian with trace equal to 1, but is allowed to
have {\em negative} eigenvalues, it is shown that $\rho_\epsilon$ can
be entangled, even for two qubits.
\end{abstract}

\pagebreak \baselineskip=4ex

The literature of Nuclear Magnetic Resonance quantum computing has
been one of the most fruitful for the past few years, since the
discovery of pseudo-pure states by Gershenfeld and Chuang
\cite{gershenfeld} and Cory et al.\cite{cory1}. Experiments
performed by different groups reported various implementations of
algorithms \cite{chuang2,jones,vandersypen}, simulations
\cite{samaroo} and quantum entanglement \cite{boulant,knill2},
including teleportation \cite{nielsen1} and other quantum effects
\cite{ollerenshaw,nelson}. These results contrast with the
landmark paper published by Braunstein and co-workers in 1999
\cite{braunstein} in which it is argued that, with the present
stage of NMR technology, {\em all} those experiments could be
interpreted through usual {\em classical correlations} between
spins. These conclusions were further extended by Linden and
Pospescu \cite{linden}. The latest of such experiments (to the
best of the authors knowledge) was performed in 2003 by Mehring
et. al \cite{mehring} who reported entanglement of nuclear and
electron spins in a molecular single-crystal, using the combined
techniques of NMR and EPR.

Reference \cite{braunstein} considered {\em
arbitrary} density matrices for $N$ qubits in the form

\begin{equation}
\rho_\epsilon = (1-\epsilon)M_d+\epsilon\rho_1 \label{eq1}
\end{equation}where $d=2^N$ is the dimension of the Hilbert space
for $N$ qubits and $M_d=I/d$ the maximally mixed density matrix.
$I$ is the identity matrix in the space of $N$ qubits, and
$\rho_1$ an {\em arbitrary density matrix}. Matrices of this form
are expanded in a basis of Pauli density matrices:

\begin{equation}
\rho = \frac{1}{2^{N}}c_{\alpha_{1}...\alpha_{N}}
\sigma_{\alpha_{1}}\otimes... \otimes\sigma_{\alpha_{N}}
\label{eq2}
\end{equation}where $\{\alpha_{s}\} \equiv \{ 0,i_{s}\}=\{0,1,2,3\}$ and
``$s$'' indicates the $s-$th qubit with the sum made over repeated
indices. Normalization imposes $c_{0\ldots 0}=1$ and the other
coefficients are in the interval $-1 \leq
c_{\alpha_{1}...\alpha_{N}} \leq 1$. From this, after a
transformation for an overcomplete basis, it is established that,
for the case $N=2$, taking the minimum value of the coefficients,
$c_{\alpha_{1},\alpha_{2}}= -1$, the bound $\epsilon\le 1/15$
limits the region below which $\rho_\epsilon$ is separable.
Generalization for arbitrary $N$ leads to $\epsilon\le 1/4^N$.
Since typically $\epsilon \approx 10^{-5}$ in NMR room temperature
liquid state experiments, the conclusion is that so far no
entanglement has ever taken place in NMR experiments, a conclusion
which has been revised by others \cite{laflamme-comment,long}. One
important observation is that in Ref. \cite{braunstein} {\em this bound is assumed to hold
independently of} $\rho_1$.

However, since the density matrix in Eq. (\ref{eq1}) is {\em
arbitrary} we can apply the reasoning to the simplest case: $N=1$
(of course, this involves no entanglement!). Taking
$c_{\alpha_{1}}=-1$, this leads to:

\begin{equation}
\rho_{N=1}=\frac12 \left( \begin{array}{cc}
0 & -1+i \\
-1-i & 2
\end{array} \right)
\label{eq3}
\end{equation}
We see that this matrix satisfies the condition
$\mbox{Tr}(\rho)=1$, but its eigenvalues are
$\lambda_1=\frac{1+\sqrt{3}}{2}$ and
$\lambda_2=\frac{1-\sqrt{3}}{2}$. Therefore, $\lambda_2<0$, and
Eq. (\ref{eq3}) {\em cannot} represent a density matrix of a
physical system \cite{despagnat}. The same can be observed for the
cases $N=2$, with $c_{\alpha_{1},\alpha_{2}}=-1$, and $N=3$, with
$c_{\alpha_{1},\alpha_{2},\alpha_{3}}=-1$, whose respective
density matrices are:

\begin{equation}
\rho_{N=2} = \frac14 \left( \begin{array}{cccc}
-2 & -2+2i & -2+2i & 2i \\
-2-2i & 2 & -2 & 0 \\
-2-2i & -2 & 2 & 0 \\
-2i & 0 & 0 & 2
\end{array} \right)
\label{eq4}
\end{equation} also satisfying  the normalization constraint $\mbox{Tr}(\rho)=1$,
but with eigenvalues $ \lambda_1=\frac{-2+2\sqrt{3}}{4},
\lambda_2=\frac{-2-2\sqrt{3}}{4}<0, \lambda_3=\lambda_4=1$, and

\[\rho_{N=3} = \frac{1}{8}\times\]
\begin{equation}
\times\left( \begin{array}{cccccccc}
-6 & -4+4i & -4+4i & 4i & -4+4i & 4i & 4i & 2+2i \\
-4-4i & 2 & -4 & 0 & -4 & 0 & -2+2i & 0 \\
-4-4i & -4 & 2 & 0 & -4 & -2+2i & 0 & 0 \\
-4i & 0 & 0 & 2 & -2-2i & 0 & 0 & 0 \\
-4-4i & -4 & -4 & -2+2i & 2 & 0 & 0 & 0 \\
-4i & 0 & -2-2i & 0 & 0 & 2 & 0 & 0 \\
-4i & -2-2i & 0 & 0 & 0 & 0 & 2 & 0 \\
2-2i & 0 & 0 & 0 & 0 & 0 & 0 & 2
\end{array} \right)
\label{eq5}
\end{equation}
with $\mbox{Tr}(\rho)=1$, and eigenvalues
$\lambda_1=\frac{-8+6\sqrt{3}}{8},
\;\;\lambda_2=\frac{-8-6\sqrt{3}}{8}<0,
\;\;\lambda_3=\lambda_4=\lambda_5= \frac{4+2\sqrt{3}}{8},
\;\;\lambda_6=\lambda_7=\lambda_8=\frac{4-2\sqrt{3}}{8}$. Since a
{\em physical} density matrix must be a positive operator
\cite{nielsen2}, none of the above matrices can be regarded as
representing a physical system.

In order to find valid intervals, consider an example where all
the coefficients are equal to some constant
$c_{\alpha_1,...,\alpha_N}=c$ in Eq.(2), and let us impose
$\lambda\ge 0$ for the eigenvalues of the resulting matrix. This
leads to the following intervals: for $N=1$, $-0.58 \leq c \leq
0.58$; for $N=2$, $-0.15\leq c\leq 0.33$; for $N=3$, $-0.05\leq
c\leq 0.15$, and so on. This case defines only one possible set of
values for the coefficients, but will be useful to derive a number
of results. Note that, for $N=2$ and
$c_{\alpha_{1},\alpha_{2}}=c$, the intervals $-1\leq c <-0.15$ and
$0.33< c \leq 1$ define infinite non-physical matrices within
$-1\leq c \leq 1$.

With the new intervals, the eigenvalues for the case $N=2$, for
$c=-0.15$, are all positive and $\rho$ satisfies
$\mbox{tr}(\rho)=1$:
\[\{\lambda_i\}=\{0.007596,0.267404,0.362499,0.362500\}\]
The same is true for $N=3$ and $c=-0.05$:
\[\{\lambda_i\}=\{0.003798,0.133702,0.165401,0.165401,0.165401,\]
\[0.122099, 0.122099, 0.122099\}\]

Note that, except for $N=1$, the intervals are {\em asymetric}.
If we write the intervals in the form $-1/A_N\le
c_{\alpha_{1}...\alpha_{N}}\le 1/B_N$ (where, for instance,
$(A_2,B_2)=(6.67,3.03),\;\; (A_3,B_3)=(20.00,6.67)$), applying the
procedure of \cite{braunstein} we find that the lower bound for
$\epsilon$ in the case $N=2$ would be $\epsilon\le A_2/15 =0.44$,
larger than the previous $\epsilon\le 1/15$ and, in the case
$N=3$, $\epsilon\le A_3/63 =0.32$, also larger than the previous
$\epsilon\le 1/64$. According to Ref. \cite{braunstein}, for
$N=2$, a matrix $\rho_{\epsilon}$ with $\epsilon>0.33$ would be
entangled, but according to the above there exist matrices in this
interval which are separable. It is important to recall that in
that reference, the bounds are independent of $\rho_{1}$. This is
not a problem of basis choice, for it also appears in the second
case considered in Ref. \cite{braunstein}, in which a continuous
overcomplete basis is used. In this case, from Eq. (9) of
\cite{braunstein} one can considers those
 $w(\vec{n}_1,\ldots,\vec{n}_N)$ which, for
 $a>1$, satisfy the relation $w(\vec{n}_1,\ldots,\vec{n}_N)\ge
-2^{2N-1}/a(4\pi)^N$, we see that $\rho_{\epsilon}$ is sepparable
for $\epsilon \le a/(a+2^{2N-1})$. In particular, for $N=2$ and
$a=6$ we see that $\rho_\epsilon$ will be separable for
$\epsilon\le 3/7$, but according to Ref. \cite{braunstein},
matrices $\rho_{\epsilon}$ with $\epsilon> 0.33$ are entangled.

Next, let us present an explicit example of separable
$\rho_\epsilon$ for $N=2$  within the entangled region given in
Ref. \cite{braunstein}. Let $\rho_{1}$ be a matrix which, in the
Pauli basis, is defined by the coefficients
$c_{\alpha_{1},\alpha_{2}}=c=-0.15$. Thus, using Eq. (\ref{eq2}),
one has:

\begin{equation}
\rho_{1}=\left( \begin{array}{cccc}
0.1375 & -0.0750+0.0750i & -0.0750+0.0750i &  0.0750i\\
-0.0750-0.0750i & 0.2875 & -0.0750 & 0 \\
-0.0750-0.0750i & -0.0750 & 0.2875 & 0 \\
-0.0750i & 0 & 0 & 0.2875
\end{array} \right)
\label{eq8}
\end{equation}This matrix satisfies $\mbox{Tr}(\rho)=1$ and has eigenvalues
$\lambda_1=0.0076, \lambda_2=0.2674, \lambda_3=\lambda_4=0.3625$.
Therefore, it is a true density matrix. Now, if we take
$\epsilon=0.40$ (that is, entangled according to Ref.
\cite{braunstein}) and replace in Eq. ({\ref{eq1}), one obtains

\begin{equation}
\rho_{\epsilon} = \left( \begin{array}{cccc}
0.2050 & -0.0300+0.0300i & -0.0300+0.0300i &  0.0300i\\
-0.0300-0.0300i & 0.2650 & -0.0300 & 0 \\
-0.0300-0.0300i & -0.0300 & 0.2650 & 0 \\
-0.0300i & 0 & 0 & 0.2650
\end{array} \right)
\label{eq9}
\end{equation}
which is also a true density matrix, since
$\mbox{Tr}(\rho_{\epsilon})=1$, and has  positive eigenvalues:
$\lambda_1=0.1530, \lambda_2=0.2569, \lambda_3=\lambda_4=0.2950$.
But, according to Peres criterium \cite{peres}, this matrix is
separable. To show that, let us expand $\rho_{\epsilon}$ in the
following basis of density matrices:

\begin{equation}
\rho_{1}=\frac12 \left( \begin{array}{cc}
1 & 0 \\
0 & 1
\end{array} \right);\;\;\;
\rho_{2}=\frac12 \left( \begin{array}{cc}
1 & -i \\
i & 1
\end{array} \right);\;\;\;
\rho_{3}=\frac12 \left( \begin{array}{cc}
1 & 1 \\
1 & 1
\end{array} \right);\;\;\;
\rho_{4}= \left( \begin{array}{cc}
1 & 0 \\
0 & 0
\end{array} \right)
\label{eq13}
\end{equation}After partial transposing \cite{peres}, one obtains
the following eigenvalues: $\lambda_1=0.1530,\lambda_2=0.2569,
\lambda_3=\lambda_4=0.2950$. Since for two qubits the Peres
criterium is a necessary and sufficient condition for separability
\cite{horodecki}, we conclude that $\rho_{\epsilon}$ is separable.
The same result can be reached using the same procedure given in
Ref.\cite{braunstein}. To see that, we need the coefficients of
$\rho_{\epsilon}$ in the Pauli basis, $d_{\alpha_{1},\alpha_{2}}$,
which are equal to $c\epsilon$ for the case the coeffcients of
$\rho_{1}$ are all equal to $c$ (see the Appendix). Replacing into
Eq. (4) of Ref. \cite{braunstein}, one has:

\[ w_{i}w_{i} + d_{i0}w_{j} + w_{i}d_{0j} + d_{ij} = \frac{1}{9} - \frac{0.15\epsilon}{3} - \frac{0.15\epsilon}{3} -0.15\epsilon >0 \]
from which one obtains the separability condition: $\epsilon \leq
0.44$, satisfied by the previous $\rho_{\epsilon}$, with
$\epsilon=0.40$. This is not the only case; there exist infinite
separable matrices $\rho_{\epsilon}$ with $\epsilon>0.33$. This
occurs because the upper and lower bounds are not fixed, as in
Ref. \cite{braunstein}, but change with $\rho_{1}$. This is why in
Eq.(9) $\epsilon>0.44$ must be considered, and not
$\epsilon>0.33$, as the correct upper bound.

Let us now build a matrix $\rho_1$ with the following set of {\em
unequal} coefficients: $c_{01}=-0.07; \;c_{02}=-0.07;\;
c_{03}=-0.07;\; c_{10}=-0.06;\; c_{11}=-0.83;\; c_{12}=-0.03;\;
c_{13}=-0.03;\; c_{20}=-0.06;\; c_{21}=-0.03;\; c_{22}=-0.03;\;
c_{23}=-0.03;\; c_{30}=-0.06;\; c_{31}=-0.03;\; c_{32}=-0.03;\;
c_{33}=-0.03$
which has eigenvalues $\{\lambda_i=0.000996, 0.078152$,
0.448038$, 0.472814\}$. Thus, for $\epsilon=0.13$ one
obtains:
\[
\rho_{\epsilon} = \left( \begin{array}{lll}
 +0.2448         & -0.0032+0.0032i    & -0.0029+0.0029i  \\
-0.0032-0.0032i & +0.2513             & -0.0279           \\
-0.0029-0.0029i & -0.0279             & +0.2506           \\
-0.0260-0.0019i & -0.0010-0.0010i     & -0.0013-0.0013i  \\
\end{array} \right.\]

\begin{equation}
\left.\begin{array}{l}
-0.0260+0.0019i \\
-0.0010+0.0010i \\
-0.0013+0.0013i \\
+0.2532         \\
\end{array}\right)
\end{equation}
with eigenvalues: $\lambda_1=0.217629$,
$\lambda_2=0.227660$, $\lambda_3=0.275745$,
$\lambda_4=0.278966$. Applying  Eq. (4) of \cite{braunstein},
one has:
\[w_{1}.w_{1} + d_{10}w_{1} + w_{1}d_{01} + d_{11}= -0.0218 < 0 \]
where $d_{i,j}$ are the coefficients of $\rho_{\epsilon}$ in the
Pauli basis (see Appendix). Therefore, $\rho_{\epsilon}$ is
entangled, but with $\epsilon$ outside the bounds established in
Ref. \cite{braunstein}. The appropriate procedure to determine
whether NMR can or cannot implement entanglement consist in finding 
the matrix $\rho_1$, for instance from quantum state tomography
experiments \cite{nielsen2}, determine the expansion coefficients
of Eq.(2) and, from those, derive the respective lower and upper
bounds which will be valid only for a particular set of matrices
$\rho_{\epsilon}$.

Finally, we will show that with the only two requirements of
$\rho_{1}$ being hermitian and satisfying $Tr(\rho_{1})=1$, a matrix
$\rho_{\epsilon}$ of two qubits can be entangled, for usual 
values of $\epsilon$. In order $\rho_{\epsilon}$ to be a true
density matrix {\em the only requirement} on $\rho_{1}$ is that it must
be hermitian and have trace equal 1. Imposing non-negative
eigenvalues (EV) for $\rho_{\epsilon}$, one has:

\begin{equation}
EV\{\rho_{\epsilon}\}=EV\{(1-\epsilon)M_{d} + \epsilon\rho_{1}\}
\geq 0    \label{eq15}
\end{equation}
Since $[M_{d}+\rho_{1}, \rho_{1}]=0$, $[M_{d}+\rho_{1}, M_{d}]=0$
and $[\rho_{1}, M_{d}]=0$, the EV of the sum is equal to the sum
of EV:

\begin{equation}
EV\{\rho_{1}\} \geq  -\frac{1-\epsilon}{\epsilon 2^{N}}
\label{eq16}
\end{equation} That is, for the usual values $\epsilon \approx 10^{-5}$, the absolute
values of the EV of $\rho_{1}$ can actually be very large.

Let us restrict ourselves to the case $N=2$ and pick the following 
particular set of values for the coefficients:
$c_{\alpha_{1}.,\alpha_{2}}=-666.66$ (note that since here
$\rho_1$ does not have to be a density matrix, these are valid
coefficients). From this, we obtain:
\begin{equation}
\rho_{1} = \left( \begin{array}{cccc}
-499.745 & -333.33+333.33i & -333.33+333.33i & 333.33i  \\
-333.33-333.33i & 166.915 & -333.33 & 0 \\
-333.33-333.33i & -333.33 & 166.915 & 0 \\
 -333.33i & 0 & 0 & 166.915
\end{array} \right)
\label{eq17}
\end{equation}
which has eigenvalues $\lambda_{1}=-1077.089496$,
$\lambda_{2}=77.599495$, $\lambda_{3}=500.244999$,
 $\lambda_{4}=500.245000$. From this, and setting $\epsilon=0.0002$
in Eq.(\ref{eq1}) we find the matrix:
\begin{equation}
\rho_{\epsilon} = \left( \begin{array}{cccc}
0.15 & -0.0666+0.0666i & -0.0666+0.0666i & 0.0667i \\
-0.0666-0.0666i & 0.2833 & -0.0666 & 0 \\
-0.0666-0.0666i & -0.0666 & 0.2833 & 0 \\
 -0.0667i & 0 & 0 & 0.2833
\end{array} \right)
\label{eq18}
\end{equation}
which has eiganvalues $\lambda_{1}=0.034532$,
$\lambda_{2}=0.265469$,
$\lambda_{3}$=0.3499989,$\lambda_{4}=0.3499990$, and therefore is
a good density matrix. According to the criterium of Ref.
\cite{braunstein} this matrix would be entangled.

Summarizing, the method of Ref. \cite{braunstein} to decide
whether NMR can or cannot produce entanglement has been revised. A
numerical analysis has shown that the interval $\pm 1$ for the
expansion coefficients is not generally correct, and that the
bounds for entanglement are dependent on the matrix $\rho_1$. The
non observation of this fact leads to the following problems: (1)
existence of density matrices with negative eigenvalues; (2)
existence of entangled matrices outside the predicted entanglement
region; (3) existence of separable matrices inside the same
entanglement region. We also show that with the requirements of
$\rho_1$ being hermitian with trace equal 1, but allowed to have
negative eigenvalues (a possible experimental situation, as
reported in Ref. \cite{laflamme}), $\rho_\epsilon$ can be
entangled, even for two-qubits. However, we were {\em not} able to
produce  entangled two qubits $\rho_\epsilon$, for $\rho_1$ being
a true density matrix (that is, a positive operator), what is in
accordance with one conclusion of Ref. \cite{braunstein}. Cases for higher numbers
of qubits must be analyzed in separate, and they may lead to further
insights into this problem.\\
\\
{\bf Acknowledgement} The authors acknowledge the support from CAPES, CNPq and FAPESP.\\
Corresponding email: jdiazb@cbpf.br 

\section{Appendix}
Any given two-qubit matrix  $\rho$  possesses the following
elements in the expansion in the Pauli basis:

\[ \begin{array}{ll}
\rho_{11}=  \frac{1}{4}(1+c_{03}+c_{30}+c_{33})        & \rho_{12}=  \frac{1}{4}(c_{01}-ic_{02}+c_{31}-ic_{32}) \\
\rho_{13}=  \frac{1}{4}(c_{10}+c_{13}-ic_{20}-ic_{23}) & \rho_{14}=  \frac{1}{4}(c_{11}-ic_{12}-ic_{21}-c_{22}) \\
\rho_{21}=  \frac{1}{4}(c_{01}+ic_{02}+c_{31}+ic_{32}) & \rho_{22}=  \frac{1}{4}(1-c_{03}+c_{30}-c_{33})        \\
\rho_{23}=  \frac{1}{4}(c_{11}+ic_{12}-ic_{21}+c_{22}) & \rho_{24}=  \frac{1}{4}(c_{10}-c_{13}-ic_{20}+ic_{23}) \\
\rho_{31}=  \frac{1}{4}(c_{10}+c_{13}+ic_{20}+ic_{23}) & \rho_{32}= \frac{1}{4}(c_{11}-ic_{12}+ic_{21}+c_{22})  \\
\rho_{33}=  \frac{1}{4}(1+c_{03}-c_{30}-c_{33})        & \rho_{34}=  \frac{1}{4}(c_{01}-ic_{02}-c_{31}+ic_{32}) \\
\rho_{41}=  \frac{1}{4}(c_{11}+ic_{12}+ic_{21}-c_{22}) & \rho_{42}=  \frac{1}{4}(c_{10}-c_{13}+ic_{20}-ic_{23}) \\
\rho_{43}=  \frac{1}{4}(c_{01}+ic_{02}-c_{31}-ic_{32}) & \rho_{44}= \frac{1}{4}( 1-c_{03}-c_{30}+c_{33}) \\
\end{array} \]

Given $\rho_1$ for $N=2$ and given $\epsilon$, the matrix
$\rho_\epsilon$ will be defined by Eq. (\ref{eq1}), which can in
turn also be expanded in the Pauli basis. Let
$d_{\alpha_{i},\alpha_{j}}$ be the expansion coefficients of
$\rho_\epsilon$. Thus, for instance, for the element $(1,1)$ we
find the following relation:

\begin{equation}
\frac{1-\epsilon}{4} + \frac{\epsilon}{4} (1+c_{03}+c_{30}+c_{33})
= \frac{1}{4}(1+d_{03}+d_{30}+d_{33}) \label{eq19}
\end{equation}
Letting $d_{\alpha_{1},\alpha_{2}}=\epsilon
c_{\alpha_{1},\alpha_{2}}$ the above equation will be valid for
any value of $\epsilon$, particularly for $c_{03}=c_{30}=c_{33}=c$
and $d_{03}=d_{30}=d_{33}=d$. It is easy to show that by setting
$c_{\alpha_{1},\alpha_{2}}=c$ e
$d_{\alpha_{1},\alpha_{2}}=d=c\epsilon$ one obtains the same
result for any two corresponding elements.

\end{document}